\documentclass[pla,reprint,onecolumn]{elsarticle}
\journal{Conference proceedings}
\usepackage{amsmath,amssymb,physics,braket,subfloat,graphicx,subfigure,epstopdf,lastpage,hyperref}
\hypersetup{colorlinks=true, urlcolor=red, citecolor=blue, linkcolor=blue}

\begin{document}
\title{Lower-vs-Higher Order Non-classicality of Photon-added Bell-type Entangled Coherent States}
\author{Deepak}
\ead{deepak20dalal19@gmail.com}

\author{Arpita Chatterjee\corref{cor1}}
\ead{arpita.sps@gmail.com}

\cortext[cor1]{Corresponding author}
\date{\today}
\address{Department of Mathematics, J. C. Bose University of Science and Technology,\\ YMCA, Faridabad 121006, India}
\begin{abstract}
We compare the lower and higher order non-classicality of a class of the photon-added Bell-type entangled coherent states (PBECS) got from Bell-type entangled coherent states using creation operators. We obtained lower and higher order criteria namely Mandel's $Q_m^l$, antibunching $d_h^{l-1}$, Subpoissioning photon statistics $D_h(l-1)$ and Squeezing $S(l)$ for the states obtained. Further we observe that first three criteria does not gives non-classicality for any state and higher order criteria gives very high positive values for all values of parameters. Also the fourth or last criterion $S(l)$ gives non-classicality for lower order as well as higher order.
\end{abstract}
\maketitle
\section{INTRODUCTION}

Since Einstein et al. \cite{y1}, who created a two-particle state that was highly entangled both in position and momentum space, entanglement has played a significant role in quantum information processing and quantum computing. The presence of a significant class of entangled states violates Bell-type inequalities \cite{y2}, which suggests that no local theory can explain their existence. Entanglement in continuous variables has been rapidly developed from both a theoretical and an experimental standpoint in the recent few years, promising to be more compact and efficient in both coding and processing quantum information \cite{y3}. A two-mode continuous-variable state is commonly referred to as an entangled coherent state (ECS) \cite{y4}. A variety of studies of ECSs have revealed their quantum nonlocality \cite{y5}. Quantum teleportation \cite{y6}, quantum computation \cite{y7}, entanglement purification \cite{y8}, quantum error corrections \cite{y9}, and other jobs have been discovered to benefit from the ECS. Furthermore, various theoretical techniques for generating ECS in cavity fields have been developed \cite{y10}.\\

A class of nonclassical states formed by stimulating specific states, on the other hand, has gotten a lot of attention. Agarwal and Tara initially proposed photon-added coherent states (PACS) \cite{y11} or excited coherent states \cite{y12}, which are the outcome of successive elementary one-photon excitations of a coherent state. In ref. \cite{y13}, the nonclassical quasi-probability distribution and amplitude of the PACS are addressed in detail. Using parametric down-conversion, the Zavatta \cite{y14} and Kalamidas \cite{y15} groups created a single PACS and a two PACS, respectively. Many authors have already presented photon-added compressed states \cite{y16}, photon-added even and (or) odd coherent states \cite{y17}, modified photon-added coherent states \cite{y18}, photon-added thermal states \cite{y19}, and so on.\\

We now introduce different class of quantum states obtained in section \ref{sec2}. Again in section \ref{sec3} we discuss the non-classicalities of different states and in the last section \ref{sec4} we gives the conclusion.

\section{Quantum States of Interest}
\label{sec2}
In this section we consider the four different states from entanglement of coherent states as discribed below;\\
\begin{align}
\label{qsi}
\ket{\psi}_1=N_1(\ket{\gamma,\gamma}+\ket{-\gamma,-\gamma}) \nonumber\\
\ket{\psi}_2=N_2(\ket{\gamma,\gamma}-\ket{-\gamma,-\gamma}) \\
\ket{\psi}_3=N_3(\ket{\gamma,-\gamma}+\ket{-\gamma,\gamma}) \nonumber\\
\ket{\psi}_4=N_4(\ket{\gamma,-\gamma}-\ket{-\gamma,\gamma}) \nonumber
\end{align}
Where $\gamma$ is the coherent state amplitude and $N_i$'s are the normalization constants for $i=1,2,3,4$ and given by \cite{y}\\
\begin{align}
N_1^{-2}=N_3^{-2}=2m!n![L_m(-|\gamma|^2)L_n(-|\gamma|^2)+e^{-4|\gamma|^2}L_m(|\gamma|^2)L_n(|\gamma|^2)] \nonumber\\
N_2^{-2}=N_4^{-2}=2m!n![L_m(-|\gamma|^2)L_n(-|\gamma|^2)-e^{-4|\gamma|^2}L_m(|\gamma|^2)L_n(|\gamma|^2)]
\end{align}
Also the expectation values of operator $a^{
\dagger p}a^q$ with respect to all the four states are given by;
\begin{equation}
\bra{\alpha,\beta}a^{p+m}b^{p+n}a^{\dagger q+m}b^{\dagger q+n}\ket{\gamma,\delta}=\left\{ 
\begin{array}{ll}
(p+m)!(p+n)!(\alpha\beta)^{* q-p}L_{p+m}^{q-p}(-\alpha^*\gamma)L_{p+n}^{q-p}(-\beta^*\delta) &\mbox{if } p\leq q\\
(q+m)!(q+n)!(\gamma\delta)^{p-q}L_{q+m}^{p-q}(-\alpha^*\gamma)L_{q+n}^{p-q}(-\beta^*\delta) &\mbox{if } q\leq p
\end{array} 
\right.
\end{equation}
\textbf{Case-I, $p\leq q$}
\begin{equation}
\label{c11}
\langle a^pa^{\dagger q}\rangle_{1,2} =N_{1,2}^2(p+m)!(p+n)!z^{*2(q-p)}2[L_{p+m}^{q-p}(-|z|^2)L_{p+n}^{q-p}(-|z|^2)\pm L_{p+m}^{q-p}(|z|^2)L_{p+n}^{q-p}(|z|^2)]
\end{equation}
\textbf{Case-II, $p\geq q$}
\begin{equation}
\label{c12}
\langle a^pa^{\dagger q}\rangle_{1,2} =N_{1,2}^2(q+m)!(q+n)!z^{2(p-q)}2[L_{q+m}^{p-q}(-|z|^2)L_{q+n}^{p-q}(-|z|^2)\pm L_{q+m}^{p-q}(|z|^2)L_{q+n}^{p-q}(|z|^2)]
\end{equation}
\textbf{Case-I, $p\leq q$}
\begin{equation}
\label{c21}
\langle a^pa^{\dagger q}\rangle_{3,4} =N_{3,4}^2(p+m)!(p+n)!z^{*2(q-p)}2(-1)^{q-p}[L_{p+m}^{q-p}(-|z|^2)L_{p+n}^{q-p}(-|z|^2)\pm L_{p+m}^{q-p}(|z|^2)L_{p+n}^{q-p}(|z|^2)]
\end{equation}
\textbf{Case-II, $p\geq q$}
\begin{equation}
\label{c22}
\langle a^pa^{\dagger q}\rangle_{3,4} =N_{3,4}^2(q+m)!(q+n)!z^{2(p-q)}2(-1)^{q-p}[L_{q+m}^{p-q}(-|z|^2)L_{q+n}^{p-q}(-|z|^2)\pm L_{q+m}^{p-q}(|z|^2)L_{q+n}^{p-q}(|z|^2)]
\end{equation}
Now it is clear from equations \eqref{c11} to \eqref{c22} that if $p=q$ then states $\ket{\psi}_3$ and $\ket{\psi}_4$ are exactly same as states $\ket{\psi}_1$ and $\ket{\psi}_2$ respectively.

\section{Non-Classicality Criteria}
\label{sec3}
In this section we find out the most important and useful criteria as (i) Mandel's $Q$ parameter(mqp), (ii) Higher and lower order antibunching $d_h^{l-1}$ (hloa), (iii) higher and lower order sub-poisoning photon statistics $D_h(l-1)$(hlsps) and (iv) higher and lower order squeezing $S(l)$(hls).
\subsection{Mandel's $Q$ Parameter}
\begin{figure}[h]
\centering
\includegraphics[scale=1]{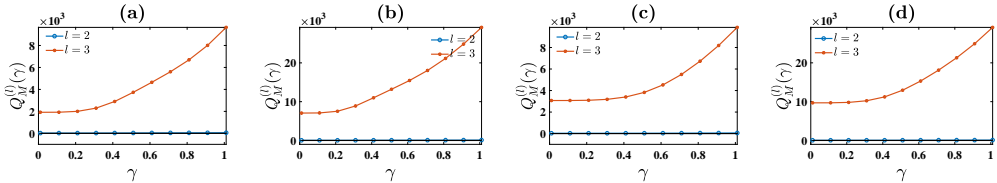}
\caption{Mandel's $Q$ as a function of $\gamma$ for lower order($l=2$) and higher order($l=3$) for $\ket{\psi}_1$ state (a) $m=1,\,n=3$, (c) $m=2,\,n=6$ and for $\ket{\psi}_2$ state (b) $m=1,\,n=3$, (d) $m=2,\,n=6$}
\label{fig1}
\end{figure}

The Mandel's parameter $Q_M$ \cite{mandel} illustrates the nonclassicality of a quantum state through its photon number distribution. The introductory definition of $Q_M$ can be generalized to an arbitrary order $l$ as \cite{sanjib}
\begin{eqnarray}
\label{eq6}
Q_M^{(l)} & = & \frac{\braket{(\Delta{\mathcal{N}})^l}}{\braket{a^\dagger a}}-1,
\end{eqnarray}
where $\Delta{\mathcal{N}}\,=\,a^\dagger a-\braket{a^\dagger a}$ is the dispersion in the number operator $\mathcal{N}=a^\dagger a$. Using the identity \cite{sanjib}
\begin{eqnarray*}
\braket{(\Delta{\mathcal{N}})^l} = \sum_{k=0}^l {l \choose k}(-1)^k\langle (a^\dagger a)^{l-k}\rangle{\langle a^\dagger a\rangle}^k
\end{eqnarray*}
and \cite{moya1}
\begin{equation}\nonumber
(a^\dagger a)^r = \sum_{n = 0}^r  S_r^{(n)}a^{\dagger n}a^n,
\end{equation}
where $S_r^{(n)}$ is the Stirling number of second kind \cite{stegun}
\begin{equation}\nonumber
S_r^{(n)} = \frac{1}{n!}\sum_{j=0}^n (-1)^{n-j}{n\choose j}j^r,
\end{equation}
the higher-order Mandel parameter $Q_M^{(l)}$ can be evaluated explicitly upto order $l$. The negativity of $Q_M^{(2)}$ signifies the negativity of the conventional Mandlel's $Q_M$. All expectations in (\ref{eq6}) have been calculated with help of (\ref{c11}) to \eqref{c22}. Also the function $Q_M^{(l)}$ as a function of $\gamma$ in fig. \ref{fig1}

Also as stated in section \ref{sec2} the behaviour of $Q_M^{(l)}$ for states $\ket{\psi}_3$ and $\ket{\psi}_4$ is same as states $\ket{\psi}_1$ and $\ket{\psi}_2$ respectively. Also from the fig. \ref{fig1} it can be notified that with increase in $m$ and $n$ the values of parameter increase for both the  states $\ket{\psi}_1$ and $\ket{\psi}_2$ and value of higher order parameter is more positive than lower order and due to positive value of $Q_M^{(l)}$ for different parameters we can't classified all the four states as they are classical or not. Further values of $Q_M^{(l)}$ is less for state $\ket{\psi}_1$ than $\ket{\psi}_2$.

\subsection{Higher and Lower-order antibunching}

\begin{figure}[h]
\centering
\includegraphics[scale=1]{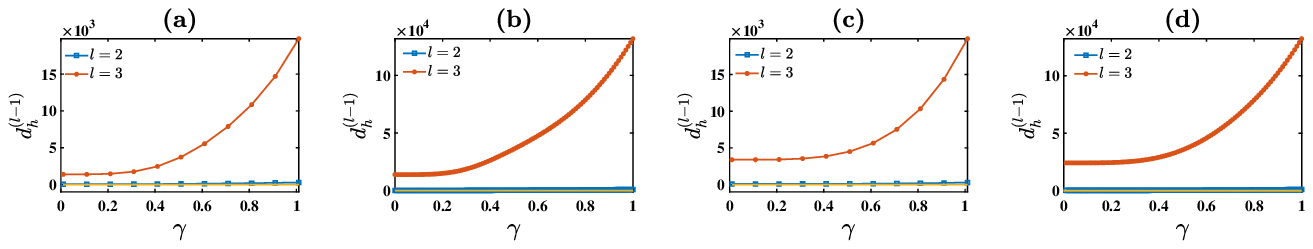}
\caption{hloa $d_h^{(l-1)}$ as a function of $\gamma$ for lower order($l=2$) and higher order($l=3$) for $\ket{\psi}_1$ state (a) $m=1,\,n=3$, (c) $m=2,\,n=6$ and for $\ket{\psi}_2$ state (b) $m=1,\,n=3$, (d) $m=2,\,n=6$}
\label{fig2}
\end{figure}

Different well-known criteria for detecting higher-order nonclassicality can be expressed in compact forms for the class of states described in \eqref{qsi}. In this subsection, we focus on higher and lower-order antibunching. The concept of hloa, by using the theory of majorization, was introduced by Lee \cite{ching}. Later it was modified by Pathak and Gracia \cite{gracia} to provide a clear physical meaning and a more simple expression. The $(l -1)$-th order antibunching is observed in a quantum state if it satisfies the following condition:
\begin{equation}
\label{eq7}
d(l-1)=\langle a^{\dagger l}a^l\rangle -{\langle a^\dagger a\rangle}^l\,\, <\,\,0
\end{equation}
Since the negativity of $d(l-1)$ indicates that the probability of photons coming bunched is less compared to that of coming independently, therefore the nonclassicality feature (\ref{eq7}) typifies how suitable the state $\ket\psi$ is as a single photon resource. Also the hloa parameter can be plotted in fig. \ref{fig2}

Also as stated in section \ref{sec2} the behaviour of $d_h^{(l-1)}$ for states $\ket{\psi}_3$ and $\ket{\psi}_4$ is same as states $\ket{\psi}_1$ and $\ket{\psi}_2$ respectively. Also from the fig. \ref{fig2} it can be notified that with increase in $m$ and $n$ the values of parameter increase for both the  states $\ket{\psi}_1$ and $\ket{\psi}_2$ and value of higher order parameter is more positive than lower order and due to positive value of $d_h^{(l-1)}$ for different parameters we can't classified all the four states as they are classical or not. Further values of $d_h^{(l-1)}$ is less for state $\ket{\psi}_1$ than $\ket{\psi}_2$.

\subsection{Higher and Lower-order sub-Poissonian photon statistics}

Higher-order sub-Poissonian photon statistics is an important feature that affirms the existence of higher-order nonclassicality of a radiation field. The lower-order antibunching and sub-Poissonian photon statistics are closely connected as the presence of later ensures the possibility of observing the first one. But recently these two phenomena are proved to be independent of each other \cite{kishore1,kishore2}. It is also reported that the higher-order antibunching and sub-Poissonian photon statistics can exist irrespective of whether their lower-order counterparts exist or not \cite{alam1}.

\begin{figure}[h]
\centering
\includegraphics[scale=1]{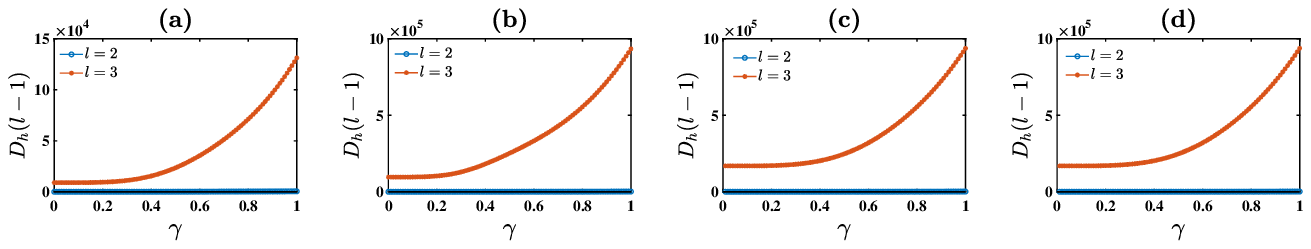}
\caption{hloa $D_h(l-1)$ as a function of $\gamma$ for lower order($l=2$) and higher order($l=3$) for $\ket{\psi}_1$ state (a) $m=1,\,n=3$, (c) $m=2,\,n=6$ and for $\ket{\psi}_2$ state (b) $m=1,\,n=3$, (d) $m=2,\,n=6$}
\label{fig3}
\end{figure}

The generalized criteria for observing the $(l-1)$-th order sub-Poissonian photon statistics (for which $\langle(\Delta\mathcal{N})^l\rangle < \langle(\Delta\mathcal{N})^l\rangle_{\ket{\mathrm{Poissonian}}}$ is given by \cite{amit}

\begin{equation}
\mathcal{D}_h(l-1) = \sum_{e=0}^{l}\sum_{f=1}^{e}S_2(e,f)^lC_e(-1)^ed(f-1){\langle a^\dagger a \rangle }^{l-e}\,\,<\,\,0
\label{eq9}
\end{equation}
where $S_2(e, f)=\sum_{r=0}^{f} {^fC_r}(-1)^r r^e$ is the Stirling number of second kind, $^lC_e$ is the usual binomial coefficient. The analytic expression of hlsps for the superposed state can be obtained by substituting (\ref{qsi}) in (\ref{eq9}). Also the obtained parameter is plotted in fig. \ref{fig3}

Also as stated in section \ref{sec2} the behaviour of $d_h^{(l-1)}$ for states $\ket{\psi}_3$ and $\ket{\psi}_4$ is same as states $\ket{\psi}_1$ and $\ket{\psi}_2$ respectively. Also from the fig. \ref{fig3} it can be notified that with increase in $m$ and $n$ the values of parameter increase for both the  states $\ket{\psi}_1$ and $\ket{\psi}_2$ and value of higher order parameter is more positive than lower order and due to positive value of $D_h^(l-1)$ for different parameters we can't classified all the four states as they are classical or not. Further values of $D_h^(l-1)$ is less for state $\ket{\psi}_1$ than $\ket{\psi}_2$.

\subsection{Higher and Lower-order squeezing}

\begin{figure}[h]
\centering
\includegraphics[scale=.6
]{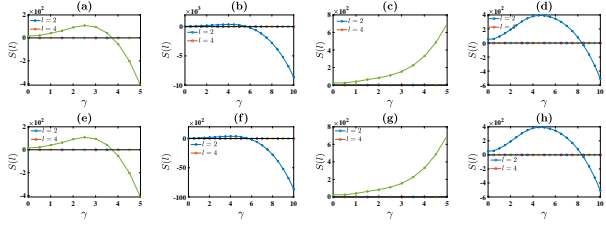}
\caption{hloa $D_h(l-1)$ as a function of $\gamma$ for lower order($l=2$) and higher order($l=3$) with $m=1,\,n=3$ for (a)$\ket{\psi}_1$, (b)$\ket{\psi}_2$, (c)$\ket{\psi}_3$, (d)$\ket{\psi}_14$ states and $m=2,\,n=6$ (a)$\ket{\psi}_1$, (b)$\ket{\psi}_2$, (c)$\ket{\psi}_3$, (d)$\ket{\psi}_4$ states}
\label{fig4}
\end{figure}

Coherent state, being the minimum uncertainty state, the product of the fluctuations in two field quadratures becomes minimum and the fluctuations in each quadrature become equal. For lower-order squeezing, the variance in one of the field quadrature (defined by a linear combination of annihilation and creation operators) reduces below the
coherent state limit at the cost of enhanced fluctuation in the other quadrature. The idea of higher-order squeezing is originated by the pioneering work of Hong and Mandel \cite{hong}. According to them, the $l$-th order higher-order squeezing ($l>2$) is obtained while the $l$-th order moment of a field quadrature operator is less than the corresponding coherent state value. Hong-Mandel's criteria for higher-order squeezing can be described by the following inequality
\begin{equation}
\label{eq10}
S(l) = \frac{\langle (\Delta X)^l \rangle -{\left(\frac{1}{2}\right)}_{\left(\frac{l}{2}\right)}}{{\left(\frac{1}{2}\right)}_{\left(\frac{l}{2}\right)}}\,\,<\,\,0,
\end{equation}
where $(x)_l$ is the conventional Pochhammer symbol and the quadrature variable is defined as $X = \frac{1}{\sqrt{2}}(a+a^\dagger)$. The inequality in (\ref{eq10}) can also be rewritten as
\begin{equation}
\label{eq11}
\langle (\Delta X)^l \rangle\,\,<\,\,{\left(\frac{1}{2}\right)}_{\left(\frac{l}{2}\right)} = \frac{1}{2^{\frac{l}{2}}}(l-1)!!,
\end{equation}
with
\begin{equation}
\begin{array}{rcl}
\label{eq12}
\langle (\Delta X)^l \rangle & = & \sum_{r=0}^{l}\sum_{i=0}^{\frac{r}{2}}\sum_{k=0}^{r-2i}(-1)^r\frac{1}{2^\frac{1}{2}}(2i-1)!^{2i}\\
& & C_k^lC_r^rC_{2i}\langle a^\dagger +a\rangle^{l-r}\langle a^{\dagger k}a^{r-2i-k}\rangle,
\end{array}
\end{equation}
where $l$ is an even number and
\begin{eqnarray*}
n!!=
\left\{
\begin{array}{lll}
& n(n-2)(n-4)\ldots 4.2 & \mbox{if $n$ is even},\\\\
& n(n-2)(n-4)\ldots3.1 & \mbox{if $n$ is odd},
\end{array}
\right.
\end{eqnarray*}
The analytic expression for the Hong-Mandel type HOS can be obtained by using (\ref{qsi}) in (\ref{eq10})-(\ref{eq12}) and it is plotted in fig \ref{fig4}.

From the fig. \ref{fig4} it can be notified that with increase in $m$ and $n$ the values of parameter increase for three states except $\ket{\psi}_3$ in both positive and negative directions and value of higher order parameter is more positive(negative) than lower order and due to negative value of $S(l)$ for different parameters we can classified  the three states except $\ket{\psi}_3$ as non-classical states. Further values of $S(l)$ for state $\ket{\psi}_3$ is positive for higher order but negative for lower order so we can conclude that higher or criterion $S(l)$ gives the better result for three states except $\ket\psi_3$ while lower order for state $\ket\psi_3$.

\section{Conclusion}
\label{sec4}

From the above discussion we can conclude that all the four states stated above are non-classical. First three criteria does not gives any result regarding the non-classical while in fourth that is in squeezing parameter three states except $\ket\psi_3$ non-classicality is identified by higher order($l=4$) parameter while for  $\ket\psi_3$ state non-classicality is identified by lower order($l=2$) criteria.


\end{document}